\documentclass[]{JHEP3}   

\JHEPspecialurl{http://jhep.sissa.it/JOURNAL/JHEP3.tar.gz}

\usepackage{epsfig}

\newcommand\fverb{\setbox\pippobox=\hbox\bgroup\verb}
\newcommand\fverbdo{\egroup\medskip\noindent%
            \fbox{\unhbox\pippobox}\ }
\newcommand\fverbit{\egroup\item[\fbox{\unhbox\pippobox}]}
\newbox\pippobox

\newcommand{\be}{\begin{eqnarray}}
\newcommand{\ee}{\end{eqnarray}}

\title{
\begin{flushright}
\normalsize{ FERMILAB-PUB-07-618-T\\FTUV 07-1105}
\end{flushright}
Self-accelerating solutions of scalar--tensor gravity}
\author{       {Gabriela Barenboim}\\
\normalsize\emph{Departament de F\'isica Te\`orica and IFIC, Universitat de
Val\`encia - CSIC}\\
\emph{Carrer Dr. Moliner 50, E-46100 Burjassot (Val\`encia), Spain}\\
Email: \email{gabriela.barenboim@uv.es}\\}

\author{\textbf{Joseph D. Lykken}\\
\normalsize\emph{Fermi National Accelerator Laboratory}\\
\emph{P.O. Box 500, Batavia, IL 60510, USA}\\
Email: \email{lykken@fnal.gov}}

\abstract{Scalar--tensor gravity is the simplest and best
understood modification of general relativity, consisting of
a real scalar field coupled directly to the Ricci scalar 
curvature. Models of this type have self-accelerating solutions. 
In an example inspired by string dilaton couplings, 
scalar--tensor gravity coupled to ordinary matter exhibits a de Sitter
type expansion, even in the presence of a {\it negative} cosmological
constant whose magnitude exceeds that of the matter density.
This unusual behavior does not require phantoms, ghosts or other
exotic sources. More generally, we show that any expansion history
can be interpreted as arising partly or entirely from scalar--tensor
gravity. To distinguish any quintessence or inflation model from
its scalar--tensor variants, we use the fact that scalar--tensor models imply
deviations of the post-Newtonian parameters of general relativity,
and time variation of the Newton's gravitational coupling $G$. 
We emphasize that next-generation probes of modified GR and the time variation of $G$
are an essential complement to dark energy probes based on luminosity-distance
measurements. 
}


\begin{document}


\section{Introduction}

There are three phenomenological approaches to explaining the
accelerating expansion of the universe.
The first is a positive cosmological constant whose magnitude
(considered as vacuum energy) somewhat 
exceeds the current matter density \cite{Weinberg:1988cp}.
The second is to introduce dynamical dark energy, usually in the form
of an ultralight quintessence scalar field \cite{Ratra:1987rm}-\cite{Copeland:2006wr}
The third is to
modify gravity so as to produce 
self-accelerating solutions \cite{Dvali:2000hr}-\cite{Carena:2006yr};
usually such solutions can be regarded as modifying the left-hand side
of the Einstein equations of motion in such a way as to simulate
a dark energy source on the right-hand side of these equations.

Scalar--tensor models \cite{Brans:1961sx}-\cite{Nordtvedt:1970uv}
can be regarded as a combination of
the latter two approaches. General relativity is modified by
a real scalar field $\theta (x)$ that couples directly to the
Ricci scalar curvature $R$. If the vacuum expectation value
of $\theta $ is dynamically evolving today, then the Einstein
equations are modified in a nonlinear way and exhibit new types
of solutions. In the absence of sources scalar--tensor models are 
classically equivalent to higher derivative modified gravity models based on a
nonlinear function $f(R)$, but this equivalence almost certainly
does not hold in realistic contexts. In fact scalar--tensor models
have a big advantage over other approaches to modified gravity,
in that it is transparent to identify regimes in these models
that are weakly--coupled and free of ghosts, violations of
the dominant energy condition, and other pathologies.

The existence of self-accelerating solutions of scalar--tensor
gravity models has already received considerable 
attention \cite{EspositoFarese:2000ij}-\cite{Demianski:2007mz}.
Strong observational constraints on such scalar--tensor cosmologies
have been derived and discussed in the 
literature \cite{Riazuelo:2001mg}-\cite{Coc:2006rt}.
Our purpose here is to exhibit some simple analytic solutions
that demonstrate the promise, weaknesses and generality of
accelerated expansion from scalar--tensor gravity. Although
we will focus on the connection to dark energy, most
of our analysis is also relevant for building models of
primordial inflation. 

In all of our examples the real scalar is ultralight.
There are two known motivations for such fields. The first
is string theory, in which the low energy effective action
can exhibit a massless dilaton and other massless moduli
fields. These generally have exponential couplings to
the Ricci scalar. They may or may not have direct couplings
to matter; when they do have such couplings, these may be
sufficiently universal to satisfy the very strong constraints
from equivalence principle tests \cite{Carroll:1998zi}.
The second motivation for
ultralight scalars (or pseudoscalars) are the pseudo-Nambu
Goldstone bosons of spontaneously broken global symmetries
that have an additional breaking due to nonperturbative
effects or to a weak explicit 
breaking \cite{Hill:1988vm}-\cite{Frieman:1995pm}.
In simple examples the
scalar potential of such PNGB's respects a discrete periodic
remnant of their original shift symmetry. We will assume
that scalar--tensor gravity implementations of this idea
imply periodic functions of $\theta$ coupling to $R$.

\section{Scalar--tensor theories}

Scalar--tensor theories are most simply defined as conventional general relativity
with a real scalar field coupled directly to the Ricci curvature. Viable models
of this type must have weak couplings between the scalar and conventional
matter or radiation. In the approximation where the scalar is decoupled from matter,
a general model can be defined in the Jordan frame:
\be\label{eqn:Jordanaction}
S &=& S_{\rm grav}(g_{\mu\nu},\theta ) + S_{\rm scalar}(g_{\mu\nu},\theta ) 
+ S_{\rm matter}(g_{\mu\nu},\psi_{\rm matter}) \; ;\\
S_{\rm grav} &=& -\frac{k^2}{4}\int d^4x\sqrt{-g} \;D(\theta )R \; ;\\
S_{\rm scalar} &=& \int d^4x\sqrt{-g}\; 
\left[ \frac{1}{2}Z(\theta )g^{\mu\nu}\partial_{\mu}\theta
\partial_{\nu}\theta - V(\theta )\right] \; .\\
\ee
where $k^2 = 1/4\pi G$, $\theta (x)$ is the scalar field rescaled by $k$
to be dimensionless, $D(\theta )$, $Z(\theta )$ and $V(\theta )$
are arbitrary functions, and $\psi_{\rm matter}$ denotes generic matter
and radiation. We use the metric signature $(+1$,$-1$,$-1$,$-1)$ and 
Wald's convention for the sign of the Riemann curvature \cite{Wald}.

\subsection{frames}
At the classical level, one is free to perform arbitrary rescalings
of the metric field and the scalar, thus obtaining many other frames
that are classically equivalent to the 
Jordan frame \cite{Capozziello:1996xg}-\cite{Faraoni:1999hp}.
For example,
$\theta$ can be redefined so as to make $Z(\theta )/k^2=1$,
giving a conventional kinetic term. However this frame is 
problematic if the original $Z(\theta )$ has zeros; note that
$Z\to 0$ is an indication that the scalar is becoming either
strongly coupled or nondynamical.

A conformal transformation of the metric can always be
found such that in the new frame $D(\theta ) = 1$. This transformation to the
Einstein frame recovers the conventional Einstein-Hilbert action,
but introduces a direct coupling between the scalar field and matter
in $S_{\rm matter}$. This transformation is also problematic if
$D(\theta )$ has zeros, an indication that gravity is becoming
strongly coupled.

By a combination of a conformal transformation and a redefinition
of the scalar, it is also possible to find a frame where
$Z(\theta ) =0$. Since the scalar is then nondynamical, it
can be eliminated by solving the constraint provided by its
equation of motion. Thus in this frame the scalar--tensor theory
becomes an $f(R)$ theory of modified general relativity. 

For most purposes the physics of scalar--tensor theories is more
transparent in the Jordan frame. Avoiding other frames also
avoids the difficult question of the status of these classical field
redefinitions in the full quantum theory. 

\subsection{equations of motion}

In the absence of matter, the equations of motion
for the general scalar--tensor theory with a Friedmann-Robertson-Walker
metric ansatz become:
\be\label{eqn:steoma}
H^2 &=& \frac{Z}{3k^2D}\dot{\theta}^2
-\frac{\dot{D}}{D}H + \frac{2}{3k^2D}V \; ;\\
\label{eqn:steomb}
\dot{H} &=& \frac{\dot{D}}{2D}H - \frac{\ddot{D}}{2D}
-\frac{Z}{k^2D}\dot{\theta}^2 \; ;\\
\label{eqn:thetaeom}
0 &=& \ddot{\theta} + 3H\dot{\theta}
+\frac{1}{2}\frac{Z'}{Z}\dot{\theta}^2
+\frac{k^2}{4}\frac{D'}{Z}R + \frac{V'}{Z} \; ,
\ee
where a prime indicates variation with respect
to the scalar field $\theta$. Here
$H = \dot{a}/a$ is the Hubble expansion rate,
and the Ricci curvature is given by
\be
R = -6(\dot{H} + 2H^2) \; .
\ee
We have assumed that the spatial curvature in the FRW
ansatz vanishes.

The third equation of motion (\ref{eqn:thetaeom}) is redundant to the
first two, which together form a coupled set of nonlinear differential
equations for $H(t)$ and $\theta (t)$.
The first equation of motion is the Friedmann equation
for scalar-tensor cosmology in the absence of matter and radiation.
Combining it with the second equation of motion, one can
derive a conventional continuity equation:
\be
\dot{\rho_{\rm eff}} + 3H(\rho_{\rm eff} + p_{\rm eff}) = 0 \; ,
\ee
where the conserved energy density $\rho_{\rm eff}$ and the corresponding
pressure $p_{\rm eff}$ are given by:
\be\label{eqn:fform}
\rho_{\rm eff} &=& \frac{3}{2}k^2H^2 \; ;\\
p_{\rm eff} &=& -\frac{1}{2}Z\dot{\theta}^2 - V - k^2\dot{H}
+\frac{3}{2}k^2\left[ H\dot{D} + H^2(D-1) \right] \; .
\ee
Of course the first relation (\ref{eqn:fform}) is a just a rewriting
of the Friedmann equation in its conventional form
$H^2 = 2\rho /3k^2$. It is important to keep in mind that
$\rho_{\rm eff}$ and $p_{\rm eff}$ differ from the
flat space energy density and pressure:
\be
\rho_{\rm eff} &=&  \rho + \Delta\rho \; ; \\
p_{\rm eff} &=& =  p + \Delta p \; ,
\ee
where
\be
\rho &=& \frac{1}{2}Z\dot{\theta}^2 + V \; ;\\
p &=& \frac{1}{2}Z\dot{\theta}^2 - V \; ;\\
\Delta\rho &=& \frac{3}{2}k^2H^2 - \frac{1}{2}Z\dot{\theta}^2 - V \; ;\\
\Delta p &=& \frac{k^2}{2}\left[ \ddot{D} + 2H\dot{D}
+(2\dot{H}+3H^2)(D-1) \right] \; .
\ee

The effective equation of state for the scalar is given by
\be
p_{\rm eff} &=& w_{\rm eff}\rho_{\rm eff} \; ;\\
w_{\rm eff} &=& -1 - \frac{2}{3}\frac{\dot{H}}{H^2} \; .
\ee
Assuming a quintessence role for the scalar $\theta$,
the effective parameter $w_{\rm eff}$ would be what is extracted,
\textit{e.g.}, from Type Ia supernovae observations.
There is no simple relation between $w_{\rm eff}$
and $w = p/\rho$.

\subsection{tracking solutions for scalar-tensor cosmologies}

The FRW solutions that are of greatest cosmological interest are
those for which the time evolution of the scalar field $\theta$
tracks the expansion rate. The simplest ansatz for this kind
of behavior is
\be\label{eqn:thetaansatz}
\dot{\theta} = -bH \; ,
\ee
where $b$ is a constant. We will restrict ourselves to solutions
of this type, although more complicated scenarios are certainly
possible.

Imposing the ansatz (\ref{eqn:thetaansatz}), the equations of
motion (\ref{eqn:steoma}-\ref{eqn:steomb}) become over-constrained.
Thus solutions of the desired type are only obtained
if the input functions $D(\theta )$, $Z(\theta )$, and
$V(\theta )$ obey a constraint, given by
\be\label{eqn:FDVconstraint}
Z = \frac{3k^2}{b^2}(D - bD'- {\mathit v}) \; ,
\ee
where we have introduced the dimensionless notation:
\be
V  \equiv \frac{3}{2}{\mathit v}k^2H^2 \; .
\ee
The equations of motion can then be solved for the effective scalar
equation of state parameter $w_{\rm eff}$:
\be\label{eqn:genwexp}
w_{\rm eff} = 1 - \frac{2}{3} \;\frac{6{\mathit v}+2bD'-b^2D''}{2D - bD'}
\; ,
\ee
in terms of which the Hubble rate is given by:
\be
H = H_0\,{\rm exp}\,\frac{1}{b}\int^{\theta} 
d\theta' \left[ 1+w(\theta' )\right]
\; .
\ee

Note that, taking $Z$, $D$ and ${\mathit v}$
as functionals of $\theta /b$, a rescaling of $b$ can be
undone by a rescaling of $\theta$ (which also implies
an overall rescaling of the kinetic function $Z$).
Thus we can take $b=1$ from now on with no loss of
generality, writing
\be\label{eqn:newfeq}
z = Z/k^2 = 3(D - D') - 2{\mathit v} \; ,
\ee
and
\be\label{eqn:genswexp}
w_{\rm eff} = 1 - \frac{2}{3} \;\frac{6{\mathit v}+2D'-D''}{2D - D'}
\; .
\ee
 
There is no known reason why the coupling functionals
$Z(\theta )$, $D(\theta )$ and $V(\theta )$ should obey the
constraint (\ref{eqn:FDVconstraint}) exactly.  However it is
certainly plausible that they satisfy this relation approximately
during a certain cosmological epoch.

\subsection{phantoms, ghosts and strong coupling}

Generic scalar--tensor lagrangians will lead to behaviors
that are unphysical, unstable or singular at the classical level,
the quantum level, or both. A partial
list of possibilities includes
\begin{itemize}
\item If $Z(\theta ) < 0$ during any epoch, the solution
has a kinetic ghost. In the cosmological context kinetic ghosts
are known as phantoms \cite{Caldwell:1999ew}-\cite{Dabrowski:2003jm}. 
They violate the weak energy condition
$p + \rho \ge 0$ and the dominant energy condition
$\rho \ge |p|$. Phantoms generically lead to singularities, 
dangerous instabilities, and pathologies in the ultraviolet
behavior of the underlying field theory \cite{Cline:2003gs,Rubakov:2006pn}. 
A successful scalar--tensor
model with a phantom epoch would have to address all of these
difficulties.
\item If $D(\theta ) < 0$ during any epoch, then the graviton
becomes a kinetic ghost. Theories of this type are believed
to be unphysical \cite{Carena:2006yr}.
\end{itemize}

Our philosophy will be to avoid such behaviors. We want
to understand the cosmological significance of scalar-tensor
theories \textit{per se}, not as examples of other exotica.

Time variation of the vacuum expectation value of $D$
corresponds to a variation in the effective strength of
the gravitational coupling. The magnitude of $\dot{D}$
is subject to strong observational bounds during certain
epochs, especially the present day \cite{Bertotti:2003rm}-\cite{Zahn:2002rr}. 
$D$ approaching zero
corresponds to gravity becoming strong.
Time variation
of the vacuum expectation value of $Z$, after a rescaling,
corresponds to changing the self-coupling of the scalar
field $\theta$, as well as its couplings to matter. The magnitude
of these latter couplings are subject to strong observational
upper bounds, at least during the present epoch. $Z$ approaching zero
corresponds to the scalar sector becoming strongly coupled,
a possibility reminiscent of the discussions in \cite{Luty}.

When we exhibit solutions that have $D \to 0$ and/or $Z\to 0$
at some time in the past, we will consider these as a 
breakdown of the modelling of the physics due to scalar-tensor
gravity sector becoming strongly coupled.

\section{De Sitter expansion with a negative cosmological constant}

Even without resorting to phantoms or other exotica, the
scalar-tensor equations of motion have many remarkable solutions.
One dramatic example is obtained by asking for a de Sitter solution,
\textit{i.e.} $H =$ constant and $w_{\rm eff}  = -1$. We will also
suppose that the scalar potential consists entirely of a cosmological
constant: ${\mathit v}(\theta ) = {\mathit v_0}$. Inserting these
ansatze into (\ref{eqn:genswexp}) yields a constraint on the
coupling $D(\theta )$:
\be
D'' - 5D' + 6(D-v_0) = 0 \; .
\ee
The solution of this constraint with the additional properties
$D(0) = 1$, $D'(0) = 0$ is given by:
\be\label{eqn:stds}
D(\theta ) &=& v_0 + 3(1-v_0){\rm e}^{2\theta} -2(1-v_0){\rm e}^{3\theta}
\; ; \\
z(\theta ) &=& 3(1-v_0)\left( -3{\rm e}^{2\theta} + 4{\rm e}^{3\theta} \right)
\; ,
\ee
where we have also assumed a form for the kinetic function satisfying
the constraint (\ref{eqn:newfeq}).

It is easily verified that the scalar--tensor theory defined by
(\ref{eqn:stds}) gives an exactly de Sitter solution $H=H_0$,
$w_{\rm eff} = -1$, for any value of the cosmological constant $v_0$,
including a negative cosmological constant. Furthermore, provided
that $v_0 < 1$, there is an epoch which includes the present
time ($\theta = 0$) where both $D(\theta )$ and $Z(\theta )$ are
positive. Thus the cosmology that we are describing does not
rely on exotic matter or ghosts. It does carry the price that
the description breaks down at some time both in the past and
in the future, where strong coupling occurs in the scalar--tensor
sector.

\begin{figure}

\centerline{\epsfxsize 3.75 truein \epsfbox {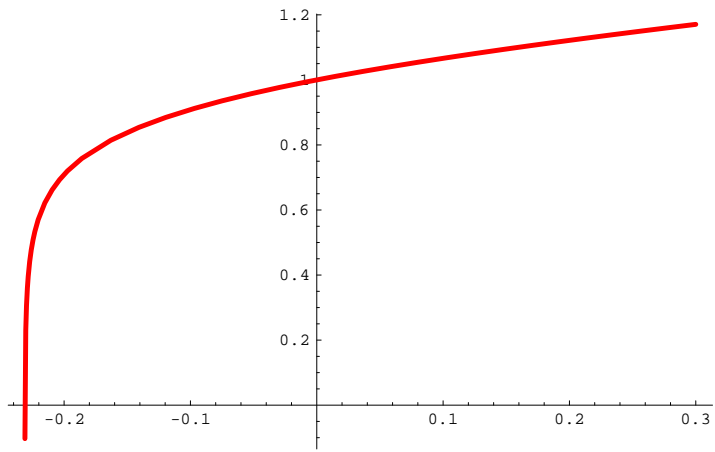} }
\caption
{\label{fig:HdsAds}
The squared Hubble rate as a function of
the scalar field expectation value $\theta$. $H$ is normalized such
that $H = H_0$ when $\theta = 0$. The scalar--tensor
theory is defined with a negative cosmological
constant $v_0 = -1$ (\textit{i.e.} a constant scalar potential
$V = -\frac{3}{2}k^2H_0^2$). Matter has been added corresponding to $\Omega_m = 0.25$.
}
\end{figure}

The solutions are even more interesting if we add conventional
matter sources. Since the scale factor $a$ is proportional
to exp$(-\theta )$, conventional matter will appear on the
right hand side of the Friedmann equation as a rescaled energy density
\be
\rho_m = \Omega_m {\rm e}^{3\theta } \; .
\ee
The exact solution to the equations of motion is
\be
\frac{H^2}{H_0^2} =
1 + \frac{\Omega_m}{(1-v_0)}\;
{\rm log}\left[ v_0 + (1-v_0){\rm e}^{3\theta } \right]
\; .
\ee
The resulting behavior is shown in Figure \ref{fig:HdsAds},
for the particular case of $\Omega_m = 0.25$ and
negative cosmological constant $v_0 = -1$. Positive
$\theta$ corresponds to the past, negative $\theta$ to
the future. In the domain plotted both $D(\theta )$ and
$Z(\theta )$ are strictly positive.

 For positive $\theta$, the Hubble rate is nearly constant,
\textit{i.e.} de Sitter-like, and slightly decreasing due to the
matter source. However in the future the negative cosmological
constant begins to dominate, and $H^2$ goes rapidly to
zero, transitioning from a nearly de Sitter metric to an anti-de
Sitter metric. Of course, a tiny negative cosmological constant
will always assert itself at some point in the future when other
sources have diluted. What is remarkable here is that the
negative cosmological constant is of the same magnitude
as the chimeric positive vacuum energy mocked up by the
effects of scalar--tensor gravity!

As expected, in the case where $v_0$ is positive and 
less than one, the expansion is de Sitter-like at $\theta = 0$
and becomes increasing de Sitter-like in the future.
In this case the effects of scalar--tensor gravity and real
positive vacuum energy conspire together. In the special
case $v_0 =1$, the solution reduces back to ordinary
inflating general relativity. 

The coupling functions $D$ and $z$ shown
in (\ref{eqn:stds}) have the form of linear combinations of
exponentials of $\theta$.  These are reminiscent of 
the effective low energy action of string theory, with
$\theta$ representing the dilaton or other related moduli
fields, and higher powers of exp$(\theta )$ representing
higher orders in the string coupling. 

Damour and Polyakov examined long ago \cite{Damour:1994zq} the
possibility of a string dilaton or similar modulus surviving as
a massless field in a phenomenolgically realistic string
compactification.  They pointed out that very stringent 
observational constraints on this scenario
can be satisified via a dynamical attraction to a local maximum
of $D(\theta )$. This is precisely what occurs in our example,
where $D$ has a local maximum at $\theta = 0$.

In order to avoid the strongest observational bounds on the time
variation of $G$, $\theta$ would have to be very close to
this local maximum today. For example the constraint from the
Cassini spacecraft \cite{Bertotti:2003rm} 
requires that $\theta \le 0.002$ today,
for the model with a negative cosmological constant $v_0 = -1$.
Even with such a tuning the model is problematical as
an explanation of the present day accelerated expansion,
since already at a redshift of 0.1 $\dot{G}/G$ is twice as
large as it is now. 

\section{PNGB gravity}

In the previous example the form of the coupling functions
was inspired by the dilaton and other moduli, the massless real scalar
fields of string theory. The other well-motivated approach to
very light scalars or pseudoscalars are pseudo-Nambu Goldstone
bosons of a spontaneously broken $U(1)$ symmetry. The PNGBs
have a shift symmetry which prevents them from appearing in the
coupling functions $D$ and $Z$ or from having a nontrivial potential $V$.
However nonperturbative effects or a weak explicit breaking can change
this picture. The simplest assumption is that $D$, $Z$ and $V$ are
restricted to be periodic functions of $\theta$, preserving a
discrete remnant $\theta \to \theta + 2\pi$ of the original
shift symmetry.

An interesting example of a scalar-tensor model of this type is
defined by
\be
D(\theta ) &=& 1 + \lambda ({\rm cos}\,\theta -1) \; ;\\
Z(\theta ) &=& \frac{1}{2}\lambda k^2 ({\rm cos}\,\theta
+{\rm sin}\,\theta ) \; ;\\
V  &=& \frac{3}{2}k^2H_0^2\left[ 1 + \lambda \left(
\frac{5}{6}({\rm cos}\,\theta + {\rm sin}\,\theta ) -1 \right)
\right]
\; .
\ee

When the dimensionless parameter $\lambda$ is taken to be small,
$\dot{G}/G$ effects are suppressed. However since the kinetic
coupling $Z$ is proportional to $\lambda$, taking $\lambda$ small
also increases the strength of $\lambda$ self-couplings as well
as any couplings of the PNGB $\theta$ to ordinary matter.
This trade-off between suppressing variations of $G$ and 
suppressing couplings to matter is a general feature of
scalar--tensor models.

In this model
the effective equation of state parameter $w_{\rm eff}$ is
exactly -1, giving a de Sitter solution. For small $\lambda$
this solution is obviously a perturbation of the standard
de Sitter solution arising from a positive cosmological constant.
The coupling $D(\theta )$ is positive for all values of $\theta$.
The kinetic coupling $Z(\theta )$ vanishes at $\theta \simeq 2.35$,
corresponding to a redshift of about 10,
indicating that the PNGB sector becomes strongly coupled.

Taking $\theta = 0$ to represent the present day, 
we note that $D(0) = 1$ and that $D$ is at a local maximum.
We can take a reasonably small value of $\lambda$,
$\lambda = 0.05$, and investigate the constraints on the model.
All of the present day bounds $\dot{G}/G$ and post-Newtonian
parameters are satisfied, provided we have tuned the present
day to coincide with $\theta = 0$ within about 3 per cent
accuracy. Furthermore in this model $\dot{G}/G$ oscillates,
so at no time in the past did $G$ differ from it's current
value by more than 10 percent.

\section{General case}

As noted in the introduction, a remarkable feature of
scalar--tensor gravity is that it allows one to obtain
\textit{any} equation of state starting from \textit{any} scalar potential,
via a suitable choice of the coupling functions $D$ and $Z$.
The major caveat is that the required $D$ and $Z$ may not be
positive, so only a subset of such models are manifestly physical.

\begin{figure}

\centerline{\epsfxsize 3.75 truein \epsfbox {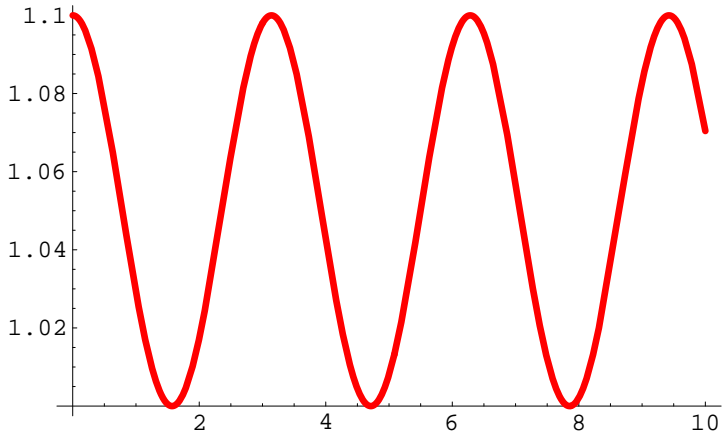} }
\caption {\label{fig:p1b}
The gravitational coupling functional $D(\theta )$ as a function of 
$\theta$.}
\end{figure}

By the same token, one can obtain any equation of state starting
from any choice of $D(\theta )$. As an example, suppose that we
want to reproduce the oscillatory equation of state of the
Slinky quintessence model \cite{Barenboim:2005np}-\cite{Barenboim:2007tu}:
\be\label{eqn:slinkyw}
w_{\rm eff} = -{\rm cos}\,2\theta \; .
\ee
At the same time, we will assume a simple oscillatory term
in the coupling $D$:
\be\label{eqn:slinkyD}
D(\theta ) = 1 + \lambda\, {\rm cos}^2\theta \; ,
\ee
where we have in mind that $\lambda$ is a small parameter.
A scalar--tensor model with the desired equation of state
is then obtained by substituting 
(\ref{eqn:slinkyD}) into (\ref{eqn:FDVconstraint}) after first
substituting (\ref{eqn:slinkyD}) and (\ref{eqn:slinkyw})
into the following expression for $v(\theta )$:
\be
v(\theta ) = \frac{1}{6}D'' - \frac{5}{6}D'
+D +\frac{1}{4}(1+w_{\rm eff})(D'-2D) \; .
\ee

\begin{figure}

\centerline{\epsfxsize 3.75 truein \epsfbox {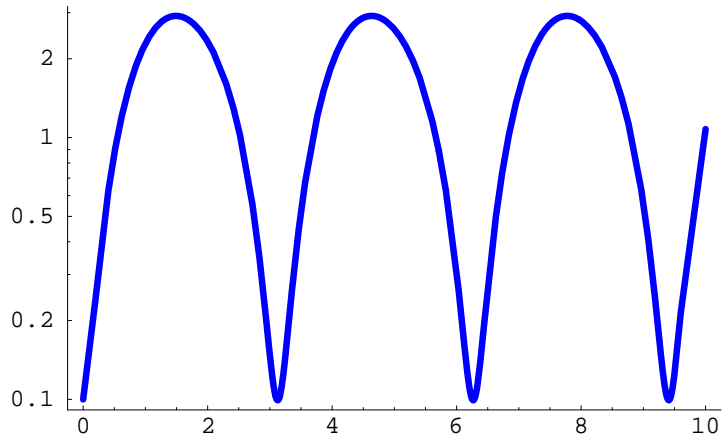} }
\caption {\label{fig:p2b}
The kinetic functional $z(\theta )$ as a function of 
$\theta$.}
\end{figure}

Taking $\lambda = 0.1$, we obtain a scalar--tensor version
of the Slinky model for which $D(\theta )$ and $Z(\theta )$
are oscillatory and strictly positive for all values
of $\theta$, as shown in Figures \ref{fig:p1b} and \ref{fig:p2b}.
The model satisfies all
the present day constraints on $\dot{G}/G$ and the post-Newtonian
parameters, provided that $\theta$ is tuned to be within about
2 per cent of a local minimum or maximum. Because $\dot{G}/G$
is oscillatory, the magnitude of $G$ never varies by more than
9 per cent from its current value.

\section{Conclusion}

We have seen that scalar-tensor gravity models can have
remarkably simple self-accelerating solutions, without resorting
to phantoms or other ghosts. These solutions are made possible
by the fact that the conserved energy associated to the scalar
is not the conventional energy that one would read off from the lagrangian
in the flat space limit. 

As we have seen, this self-accelerating
feature can even overcome, for a time, the effects of a negative
cosmological constant of similar magnitude. In such a scenario,
our immediate cosmological future exactly contradicts what one
would predict from a naive extrapolation of the current expansion.
Although the model we exhibited was not entirely realistic,
it has an intriguing connection to previous attempts to 
construct realistic string models with ultralight moduli. 

Scalar-tensor models are very constrained by observational data.
It does not appear likely that one could account for
dark energy entirely from the effects of scalar-tensor gravity,
especially in a framework where the gravity sector is manifestly
weakly coupled and ghost--free.

However we gave two examples of realistic models where the novel
properties of scalar-tensor gravity play an important role
in the current accelerated expansion. These models are somewhat
tuned in order to satisfy present day limits on the time variation
of $G$. They predict interesting variations of $G$, on the order
or 10\% , at earlier times. Improved CMB \cite{Zahn:2002rr}
or neutron star \cite{Thorsett:1996fr} constraints would directly
test such scenarios, as would galaxy cluster based tests of
modified gravity \cite{Zhang:2007nk,Schmidt:2007vj}.

More generally, our examples show the importance of using
a broad-based observational approach to dark energy.
The ambitious Stage IV dark energy probes currently
planned \cite{Albrecht:2006um} are certainly not sufficient
by themselves \cite{Boisseau:2000pr,Capozziello:2007iu}
to disentangle scalar-tensor effects from
quintessence and other scenarios. 

The self-accelerating properties of scalar-tensor models
look promising for models of primordial inflation.
It appears that such models could have
virtues similar to models of hybrid inflation \cite{GarciaBellido:1995fz,Linde:1993cn}.
Variations of $G$ at times prior to Big Bang Nucleosynthesis are
hardly constrained. In this arena it also seems more promising
to forge a concrete link
between scalar-tensor gravity and string theory.

\subsection*{Acknowledgments}
The authors are grateful to Chris Hill
and Jos\'e Santiago for useful comments. 
GB acknowledges support from the Spanish MEC and FEDER under 
Contract FPA 2005/1678,
and the Generalitat Valenciana under Contract GV05/267.
Fermilab is operated by the Fermi Research Alliance LLC under
contact DE-AC02-07CH11359 with the U.S. Dept. of Energy.

\newpage



\begin{thebibliography}{99}



\bibitem{Weinberg:1988cp}
  S.~Weinberg,
  Rev.\ Mod.\ Phys.\  {\bf 61}, 1 (1989).


\bibitem{Ratra:1987rm}
  B.~Ratra and P.~J.~E.~Peebles,
  Phys.\ Rev.\  D {\bf 37}, 3406 (1988).

\bibitem{Peebles:1987ek}
  P.~J.~E.~Peebles and B.~Ratra,
  Astrophys.\ J.\  {\bf 325}, L17 (1988).

\bibitem{Carroll:1998zi}  
S.~M.~Carroll,  
Phys.\ Rev.\ Lett.\  {\bf 81}, 3067 (1998)  
[arXiv:astro-ph/9806099]. 

\bibitem{Copeland:2006wr}
  E.~J.~Copeland, M.~Sami and S.~Tsujikawa,
  Int.\ J.\ Mod.\ Phys.\  D {\bf 15}, 1753 (2006)
  [arXiv:hep-th/0603057].


\bibitem{Dvali:2000hr}
  G.~R.~Dvali, G.~Gabadadze and M.~Porrati,
  Phys.\ Lett.\  B {\bf 485}, 208 (2000)
  [arXiv:hep-th/0005016].

\bibitem{Carroll:2003wy}
  S.~M.~Carroll, V.~Duvvuri, M.~Trodden and M.~S.~Turner,
  Phys.\ Rev.\  D {\bf 70}, 043528 (2004)
  [arXiv:astro-ph/0306438].

\bibitem{Dolgov:2003px}
  A.~D.~Dolgov and M.~Kawasaki,
  Phys.\ Lett.\  B {\bf 573}, 1 (2003)
  [arXiv:astro-ph/0307285].

\bibitem{Nojiri:2003ft}
  S.~Nojiri and S.~D.~Odintsov,
  Phys.\ Rev.\  D {\bf 68}, 123512 (2003)
  [arXiv:hep-th/0307288].

 \bibitem{Mena:2005ta}
  O.~Mena, J.~Santiago and J.~Weller,
  Phys.\ Rev.\ Lett.\  {\bf 96}, 041103 (2006)
  [arXiv:astro-ph/0510453].

\bibitem{Nojiri:2006ri}
  S.~Nojiri and S.~D.~Odintsov,
  Int.\ J.\ Geom.\ Meth.\ Mod.\ Phys.\  {\bf 4}, 115 (2007)
  [arXiv:hep-th/0601213].

\bibitem{Carena:2006yr}
  M.~S.~Carena, J.~Lykken, M.~Park and J.~Santiago,
  Phys.\ Rev.\  D {\bf 75}, 026009 (2007)
  [arXiv:hep-th/0611157].


\bibitem{Brans:1961sx}
  C.~Brans and R.~H.~Dicke,
  Phys.\ Rev.\  {\bf 124}, 925 (1961).

\bibitem{Bergmann:1968ve}
  P.~G.~Bergmann,
  Int.\ J.\ Theor.\ Phys.\  {\bf 1}, 25 (1968).

\bibitem{Wagoner:1970vr}
  R.~V.~Wagoner,
  Phys.\ Rev.\  D {\bf 1}, 3209 (1970).

\bibitem{Nordtvedt:1970uv}
  K.~J.~Nordtvedt,
  Astrophys.\ J.\  {\bf 161}, 1059 (1970).



\bibitem{EspositoFarese:2000ij}
  G.~Esposito-Farese and D.~Polarski,
  Phys.\ Rev.\  D {\bf 63}, 063504 (2001)
  [arXiv:gr-qc/0009034].

\bibitem{Faraoni:2004dn}
  V.~Faraoni,
  Phys.\ Rev.\  D {\bf 70}, 044037 (2004)
  [arXiv:gr-qc/0407021].

\bibitem{Carvalho:2004ty}
  F.~C.~Carvalho and A.~Saa,
  Phys.\ Rev.\  D {\bf 70}, 087302 (2004)
  [arXiv:astro-ph/0408013].

\bibitem{Nojiri:2005pu}
  S.~Nojiri and S.~D.~Odintsov,
  Gen.\ Rel.\ Grav.\  {\bf 38}, 1285 (2006)
  [arXiv:hep-th/0506212].

\bibitem{Demianski:2007mz}
  M.~Demianski, E.~Piedipalumbo, C.~Rubano and P.~Scudellaro,
  arXiv:0711.1043 [astro-ph].



\bibitem{Riazuelo:2001mg}
  A.~Riazuelo and J.~P.~Uzan,
  Phys.\ Rev.\  D {\bf 66}, 023525 (2002)
  [arXiv:astro-ph/0107386].

\bibitem{Catena:2004ba}
  R.~Catena, N.~Fornengo, A.~Masiero, M.~Pietroni and F.~Rosati,
  Phys.\ Rev.\  D {\bf 70}, 063519 (2004)
  [arXiv:astro-ph/0403614].

\bibitem{Carroll:2004hc}
  S.~M.~Carroll, A.~De Felice and M.~Trodden,
  Phys.\ Rev.\  D {\bf 71}, 023525 (2005)
  [arXiv:astro-ph/0408081].

\bibitem{Gannouji:2006jm}
  R.~Gannouji, D.~Polarski, A.~Ranquet and A.~A.~Starobinsky,
  JCAP {\bf 0609}, 016 (2006)
  [arXiv:astro-ph/0606287].

\bibitem{Coc:2006rt}
  A.~Coc, K.~A.~Olive, J.~P.~Uzan and E.~Vangioni,
  Phys.\ Rev.\  D {\bf 73}, 083525 (2006)
  [arXiv:astro-ph/0601299].


\bibitem{Hill:1988vm}
  C.~T.~Hill, D.~N.~Schramm and J.~N.~Fry,
  Comments Nucl.\ Part.\ Phys.\  {\bf 19}, 25 (1989).

\bibitem{Freese:1990rb}
  K.~Freese, J.~A.~Frieman and A.~V.~Olinto,
  Phys.\ Rev.\ Lett.\  {\bf 65}, 3233 (1990).

\bibitem{Frieman:1991tu}  
J.~A.~Frieman, C.~T.~Hill and R.~Watkins, 
Phys.\ Rev.\  D {\bf 46}, 1226 (1992).  

\bibitem{Frieman:1995pm}
  J.~A.~Frieman, C.~T.~Hill, A.~Stebbins and I.~Waga,  
Phys.\ Rev.\ Lett.\  {\bf 75}, 2077 (1995)  
[arXiv:astro-ph/9505060].  


\bibitem{Wald}
R. Wald, {\sl ``General Relativity''}, University of Chicago Press,
1984.




\bibitem{Capozziello:1996xg}
  S.~Capozziello, R.~de Ritis and A.~A.~Marino,
  Class.\ Quant.\ Grav.\  {\bf 14}, 3243 (1997)
  [arXiv:gr-qc/9612053].

\bibitem{Dick:1998ke}
  R.~Dick,
  Gen.\ Rel.\ Grav.\  {\bf 30}, 435 (1998).

\bibitem{Faraoni:1999hp}
  V.~Faraoni and E.~Gunzig,
  Int.\ J.\ Theor.\ Phys.\  {\bf 38}, 217 (1999)
  [arXiv:astro-ph/9910176].


\bibitem{Caldwell:1999ew}
  R.~R.~Caldwell,
  Phys.\ Lett.\  B {\bf 545}, 23 (2002)
  [arXiv:astro-ph/9908168].

\bibitem{Gibbons:2003yj}
  G.~W.~Gibbons,
  arXiv:hep-th/0302199.

\bibitem{Nojiri:2003vn}
  S.~Nojiri and S.~D.~Odintsov,
  Phys.\ Lett.\  B {\bf 562}, 147 (2003)
  [arXiv:hep-th/0303117].

\bibitem{Dabrowski:2003jm}
  M.~P.~Dabrowski, T.~Stachowiak and M.~Szydlowski,
  Phys.\ Rev.\  D {\bf 68}, 103519 (2003)
  [arXiv:hep-th/0307128].


\bibitem{Cline:2003gs}
  J.~M.~Cline, S.~Jeon and G.~D.~Moore,
  Phys.\ Rev.\  D {\bf 70}, 043543 (2004)
  [arXiv:hep-ph/0311312].

\bibitem{Rubakov:2006pn}
  V.~A.~Rubakov,
  Theor.\ Math.\ Phys.\  {\bf 149}, 1651 (2006)
  [Teor.\ Mat.\ Fiz.\  {\bf 149}, 409 (2006)]
  [arXiv:hep-th/0604153].




\bibitem{Bertotti:2003rm}
  B.~Bertotti, L.~Iess and P.~Tortora,
  Nature {\bf 425}, 374 (2003).


\bibitem{Williams:1995nq}
  J.~G.~Williams, X.~X.~Newhall and J.~O.~Dickey,
  Phys.\ Rev.\  D {\bf 53}, 6730 (1996).


\bibitem{Chiba:2001ui}
  T.~Chiba,
  arXiv:gr-qc/0110118.


\bibitem{Thorsett:1996fr}
  S.~E.~Thorsett,
  Phys.\ Rev.\ Lett.\  {\bf 77}, 1432 (1996)
  [arXiv:astro-ph/9607003].


\bibitem{Zahn:2002rr}
  O.~Zahn and M.~Zaldarriaga,
  Phys.\ Rev.\  D {\bf 67}, 063002 (2003)
  [arXiv:astro-ph/0212360].



\bibitem{Luty}
  M.~A.~Luty, M.~Porrati and R.~Rattazzi,
  JHEP {\bf 0309}, 029 (2003)
  [arXiv:hep-th/0303116].


\bibitem{Damour:1994zq}
  T.~Damour and A.~M.~Polyakov,
  Nucl.\ Phys.\  B {\bf 423}, 532 (1994)
  [arXiv:hep-th/9401069].


\bibitem{Barenboim:2005np} 
G.~Barenboim and J.~D.~Lykken,  
Phys.\ Lett.\  B {\bf 633}, 453 (2006)  
[arXiv:astro-ph/0504090]. 

\bibitem{Barenboim:2006rx}  
G.~Barenboim and J.~D.~Lykken,  
JHEP {\bf 0607}, 016 (2006) 
[arXiv:astro-ph/0604528].  

\bibitem{Barenboim:2006jh}  
G.~Barenboim and J.~D.~Lykken,  
JHEP {\bf 0612}, 005 (2006)  
[arXiv:hep-ph/0608265]. 

\bibitem{Barenboim:2007tu}
  G.~Barenboim and J.~D.~Lykken,
  arXiv:0707.3999 [astro-ph].


\bibitem{Zhang:2007nk}
  P.~Zhang, aff, R.~Bean and S.~Dodelson,
  arXiv:0704.1932 [astro-ph].

\bibitem{Schmidt:2007vj}
  F.~Schmidt, M.~Liguori and S.~Dodelson,
  Phys.\ Rev.\  D {\bf 76}, 083518 (2007)
  [arXiv:0706.1775 [astro-ph]].


\bibitem{Albrecht:2006um}
  A.~Albrecht {\it et al.},
  arXiv:astro-ph/0609591.



\bibitem{Boisseau:2000pr}
  B.~Boisseau, G.~Esposito-Farese, D.~Polarski and A.~A.~Starobinsky,
  Phys.\ Rev.\ Lett.\  {\bf 85}, 2236 (2000)
  [arXiv:gr-qc/0001066].

\bibitem{Capozziello:2007iu}
  S.~Capozziello, S.~Nesseris and L.~Perivolaropoulos,
  arXiv:0705.3586 [astro-ph].


\bibitem{GarciaBellido:1995fz}
  J.~Garcia-Bellido and D.~Wands,
  Phys.\ Rev.\  D {\bf 52}, 6739 (1995)
  [arXiv:gr-qc/9506050].

\bibitem{Linde:1993cn}
  A.~D.~Linde,
  Phys.\ Rev.\  D {\bf 49}, 748 (1994)
  [arXiv:astro-ph/9307002].


\end{thebibliography}
\end{document}